\documentclass[12pt]{article}

\usepackage{amsmath, amssymb}

\usepackage{graphicx}
\usepackage{color}
\definecolor{rltbrightred}{rgb}{1,0,0}
\definecolor{rltred}{rgb}{0.75,0,0}
\definecolor{rltdarkred}{rgb}{0.5,0,0}
\definecolor{rltbrightgreen}{rgb}{0,0.75,0}
\definecolor{rltgreen}{rgb}{0,0.5,0}
\definecolor{rltdarkgreen}{rgb}{0,0,0.25}
\definecolor{rltbrightblue}{rgb}{0,0,1}
\definecolor{rltblue}{rgb}{0,0,0.75}
\definecolor{rltdarkblue}{rgb}{0,0,0.5}

\definecolor{webred}{rgb}{0.5,.25,0}
\definecolor{webblue}{rgb}{0,0,0.75}
\definecolor{webgreen}{rgb}{0,0.5,0}

\definecolor{webdarkblue}{rgb}{0,0,0.5}
\definecolor{webbrightgreen}{rgb}{0,0.75,0}

\newcommand{\mydef}
        {\stackrel{\mathrm{def}}{=}}
\newcommand{\units}[1]
	{\mathrm{\, \scriptstyle #1}}

\newcommand{\CS}{\ensuremath{\mathcal{S}}}
\newcommand{\CI}{\ensuremath{\mathcal{I}}}
\newcommand{\CR}{\ensuremath{\mathcal{R}}}

\newcommand{\etal}{{\it et al.\ }}
\newcommand{\ie}{{\it i.e.\ }}
\newcommand{\eg}{{\it e.g.\ }}

\title{Equation-free modeling of evolving diseases: 
Coarse-grained computations with individual-based models.}

\author{Jaime Cisternas${}^{1,2}$,
C. William Gear${}^{2,3}$,\\
Simon Levin${}^4$
and Ioannis G. Kevrekidis${}^{1,2}$
\footnote{
Corresponding author, {\tt yannis@princeton.edu}} \\
\small ${}^1$ Program in Applied and Computational Mathematics,
Princeton University, Princeton, NJ 08544; \\
\small ${}^2$ Department of Chemical Engineering,
Princeton University, Princeton, NJ 08544; \\
\small ${}^3$ NEC Research Institute, Princeton, NJ 08540; and \\
\small ${}^4$ Department of Ecology and Evolutionary Biology,
Princeton University, Princeton, NJ 08544.}

\date{\today}

\begin{document}\label{start}

\maketitle



\abstract{
We demonstrate how direct simulation of stochastic, individual-based models
can be combined with continuum numerical analysis techniques to 
study the dynamics of evolving diseases.  
Sidestepping the necessity of obtaining explicit population-level models, 
the approach analyzes the (unavailable in closed form)
`coarse' macroscopic equations, estimating the necessary 
quantities through appropriately initialized, short `bursts'
of individual-based dynamic simulation.  
We illustrate this approach by analyzing a stochastic and discrete model for the
evolution of disease agents caused by point mutations within individual hosts.
Building up from classical SIR and SIRS models, our example uses a one-dimensional 
lattice for variant space, and assumes a finite number of individuals.
Macroscopic computational tasks enabled through this approach include
stationary state computation, coarse projective integration, parametric 
continuation and stability analysis.}


\section{Introduction}
Classical models of disease dynamics rely on systems of
differential equations that represent the numbers of
individuals in various categories through continuous
variables, allowing for infinitesimal population densities.
This is especially problematic in examining the evolution
of disease agents, since it ignores the stochastic nature
of the dynamics of rare mutants.
Using an equation-free multiscale computational approach
may circumvent this problem, enabling individual-based stochastic simulations
to analyze the macroscopic, {\it expected} epidemic dynamics directly.
Assuming that continuum, population level evolution equations for the coarse-grained epidemic
dynamics exist (even though they are not available in closed form), we 
circumvent their explicit derivation using the so-called coarse time-stepper
Refs.~\cite{Coarse,GKT02,Manifesto}.

Traditional continuum numerical analysis codes typically operate by repeatedly 
calling -with the current state and parameters as input- a subroutine that contains the model,
usually in the form of the time derivative of the population-level equations.
The subroutine {\it evaluates} the model and returns the time derivative; possibly also
partial derivatives with respect to the state variables (Jacobians) or with respect to 
parameters are returned.
The main code uses the results of these {\it function evaluations} to perform the
computations underlying computational tasks such as numerical integration, 
fixed point computation (\eg Newton-Raphson iteration) to locate steady states, stability,
continuation and bifurcation computations.
When explicit coarse-grained equations are available, the same type of repeated function 
evaluations underpin computational tasks such as optimization, controller design or 
the computation of coarse self-similar solutions.

In evolutionary epidemiology computations (as is often the case in contemporary modeling in 
the physical, life and social sciences) good system descriptions are available at a fine-scale,
stochastic, individual-based level.
The closures required to translate the individual-based dynamics to macroscopic equations,
amenable to systematic computer-aided analysis (such as bifurcation, stability, continuation  
and parametric analysis) are unavailable in closed form.
Our equation-free multiscale computational framework replaces function evaluations (which we
would perform if we had explicit macroscopic equations) with short bursts of appropriately initialized
ensembles of simulations with the fine-scale, individual based model. 
The results of these short simulations are used to {\it estimate} the same quantities
(time derivatives, residuals, the action of Jacobians) that would be {\it evaluated} from
a population-level model.
The macroscopic tasks proceed in the same way, and with the same algorithms as before; the
only difference is that function evaluation has been substituted by short bursts of
microscopic/stochastic simulation and appropriate processing of the results (using 
established system identification techniques).

This two-level (possibly multi-level) ``closure on demand" framework constitutes
a bridge between simulation at a fine level (atomistic/ individual based / stochastic) 
and exploration of emergent behavior at a coarser level (macroscopic / averaged / expected).
Remarkably, a large body of work on the so-called matrix-free large scale scientific
computation methods over the last twenty years fits directly in our multiscale framework; hence the
``equation-free" characterization.
The approach, first proposed in Ref.~\cite{Coarse} and discussed
in detail in \cite{Manifesto}, has been validated in a number
of contexts (coarse kinetic Monte Carlo simulations of
lattice gas models of surface reactions, coarse Brownian dynamics
for nematic liquid crystals, coarse molecular dynamics of
peptide fragments, and coarse Lattice Boltzmann simulations
of multiphase flow, to name a few).
A discussion of the approach, containing references
to related approaches, such as the optimal predictors
of Chorin and coworkers \cite{Chorin98,Chorin00},
as well as the Quasi-Continuum
method of Phillips, Ortiz and coworkers \cite{Phillips99} can be found
in Ref.~\cite{Manifesto}; for relations with accelerated
molecular dynamics techniques see \cite{Hummer}.
This is our first attempt at using this approach in an epidemiological
(and in particular in an evolutionary, mutation-based) context.
Coarse integration schemes, in our illustrative example,
will accelerate the direct 
simulation of the  expected epidemic dynamics.
Fixed point algorithms will be used to converge on stationary drifting
population ``pulse" shapes, their expected drift speeds, and their dependence
on model parameters, including the overall population size.

To motivate our SIRS-type evolutionary illustrative example, it is interesting 
to draw some parallels with the special case of the evolution of the influenza virus.
Influenza is one of the most familiar diseases humans face,
having coexisted with us for perhaps four centuries.  
Despite this,
and despite the fact that recovery from a particular strain of flu
confers lifetime resistance, the disease remains a scourge, killing
millions annually (Earn \etal \cite{Earn}). 
The key reason lies in the evolutionary dynamics of the virus.

Study of the evolution of the influenza virus has been greatly
enhanced in recent years by advances in molecular biology, and
has helped elucidate the unusual phylogenetic patterns that emerge.
In particular (Fitch \etal \cite{Fitch00}), the evolutionary tree of the most
common (H3N2) subtype of the influenza A virus has the form of a
single major trunk, with limited variation about it.
This makes it
possible to study the evolution of the virus in terms of movement
along a one-dimensional continuum, as presented by Lin \etal \cite{LACL}.

In this work we want to study evolutionary epidemiological models
without making the assumptions of an infinite population and a
continuous space of variants. 
The total population is taken to consist of a finite number 
of individuals (of the order of $10^2$--$10^7$),
and the space of variants is a discrete, one-dimensional lattice. 
The apparent asymptotic behavior of the stochastic
model consists of
effectively stationary shape population ``pulses", drifting at an 
essentially constant rate over variant space.
The paper is organized as follows: in the next Section our stochastic,
individual-based model will be presented. Coarse computational methods
are described in Section 3, and the results of their application 
constitute Section 4. We conclude with a brief discussion in Section 5.

\begin{figure}[t!]
\begin{center}
\resizebox{10cm}{!}{
\includegraphics[angle=270]{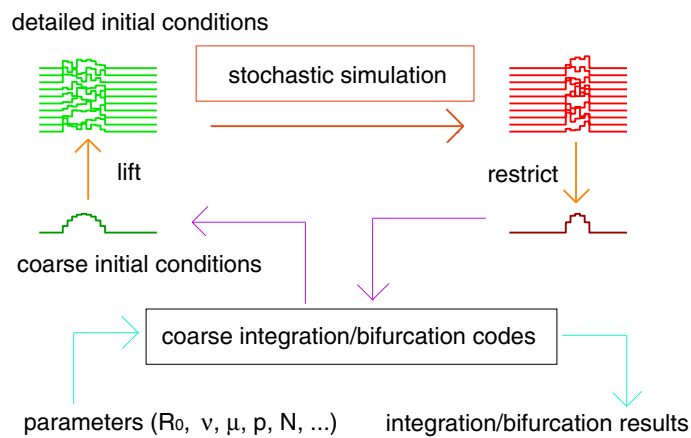}
}
\end{center}
\caption[]{Schematic of the coarse time-stepper, employing
a low-dimensional representation for the system statistics,
lifting and restriction operators, and the individual-based, 
stochastic simulation code.
Evaluating this time-stepper for appropriate sets of initial
conditions helps estimate coarse-grained time derivatives, 
enables projective integration methods, as well as coarse fixed point
computation/continuation through matrix-free iterative algorithms.}
\label{diagram}
\end{figure}

\section{The Stochastic Model}


Our SIRS-type illustrative model is partially based on certain features of 
Influenza A virus evolution.
Following an approach of Pease \cite{Pease87},  Andreasen \etal \cite{ALL}
developed a continuum model for the evolution of a virus.  
In particular,
they characterize individuals according to the last strain of the virus
they experienced, since this is the best indicator of current susceptibility.
For influenza A, recovery from (or vaccination against) a particular
strain of the virus
confers lifetime immunity not only to that strain, but also (at least
partially) to the closely related strains that appear in subsequent years.
%
%
Based on this, they demonstrate that asymptotic traveling waves form
in a continuum model, and
move with a characteristic speed through sequence space - a speed that can
be compared with measured speeds of ``drift'' evolution (the gradual change
in the virus due to mutation and selection) \cite{LACL}.
Earn \etal \cite{Earn}, echoing Andreasen \etal \cite{ALL} and Lin \etal
\cite{LACL}, note that the continuum model is a gross approximation, and
that more accuracy would be possible with an individual-based model, but
at the cost of analytical tractability.  
Our coarse-grained approach aims at circumventing the analytical tractability issue;
we solve the analytically unavailable equations by acting directly, through appropriately
initialized calls, to an individual-based simulator.
 
Our work is based upon classical SIRS-type  models of an epidemic.
For influenza A, the appropriate model would be SIR since
immunity in this case is permanent (Refs.~\cite{LACL,ALL}).
In a closed population, each individual will be either susceptible ($\CS$),
infected ($\CI$) or recovered ($\CR$),
meaning that he/she has not suffered the disease,
is currently a carrier of the virus,
or had the disease in the past and has immunity to a future infection.
Other details like age, gender or time of residence in
each category are not considered here.
No interaction structure is assumed, so we
use the equivalent of random mixing.

Our model, based on a one-dimensional lattice of virus strains, is a compartmental one.
We use the variable $S \mydef [\CS] \in \mathbb{Z}^+_0$ for
the number of individuals in compartment $\CS$.
Similarly for the other compartments we define the variables
$I \mydef [\CI]$ and $R \mydef [\CR]$.
Possible changes of state (transition between compartments)
are infections, recoveries, losses of immunity, births and deaths.
The SIRS model
also includes the loss of immunity after certain period of time after
the recovery (not the case in influenza A).
As a result of errors in the replication of virions within the hosts,
new variants of the virus are created. 
These new strains may be able to overcome the defenses of individuals
who had recovered from a previous epidemic.

We designate a single compartment $\CS$ for totally naive individuals,
and subdivide $\CI$ and $\CR$ into sub-compartments:
$\CI_i$ individuals who are infected with the strain $i$, and
$\CR_i$ individuals who have recovered from the strain $i$.
Though all strains are assumed equal
(in their dynamic rates), cross-immunity
induces a hierarchy that explains a preferred direction
in the time evolution of the epidemic outburst:
the fact that an individual in $\CR_i$ is immune to the disease carried
by the $\CI_j$, does not mean that individuals in $\CR_j$ are immune to
$\CI_i$. 
In Table~\ref{processes} a summary of the transitions between
compartments is presented, along with their transition rates.
All the processes are assumed Poisson.
We denote $I \mydef \Sigma_i I_i$, $R \mydef \Sigma_i R_i$ and
$N \mydef S+I+R$.
The population size is not constant but for small $\mu$ the variation
is small over finite time computational horizons.

\begin{table}[t!]
\begin{center}
\caption[]{Allowed transitions between different compartments and
their corresponding probability rates per individual.
Here $\mathcal{X}$ means any $\CS, \CI_i$ or $\CR_i$ individual,
and $\emptyset$ means a death.
For influenza A, immunity to a particular strain is lifelong,
so $\gamma=0$. Subscripts refer to particular strains.
$K_{i,j}$ is the relative susceptibility of an $\CR_j$ individual
to infection with strain $i$.}
\begin{tabular}{lccc}
\hline
transition & from & to & rate \\
\hline
infections   & $\CS$ & $\CI_i$ & $\beta I_i / N$ \\
reinfections & $\CR_j$ & $\CI_i$ & $\beta K_{i,j} I_i /N$ \\
recoveries   & $\CI_i$ & $\CR_i$ & $\nu$ \\
births       & $\mathcal{X}$ & $\mathcal{X} + \CS$ & $\mu$ \\
deaths       & $\CS$ & $\emptyset$ & $\mu$ \\
             & $\CI_i$ & $\emptyset$ & $\mu$ \\
             & $\CR_i$ & $\emptyset$ & $\mu$ \\
losses of immunity & $\CR_i$ & $\CS$ & $\gamma$ \\
mutations    & $\CI_i$ & $\CI_{i-1}$ & $p/2$ \\
             & $\CI_i$ & $\CI_{i+1}$ & $p/2$ \\
\hline
\end{tabular}

\label{processes}
\end{center}
\end{table}


The epidemiological parameters are the transition rates:
$\nu, \mu, \gamma, \beta$ and $p$ (in units of $1/$year),
but it is convenient to introduce the basic reproduction number
(a nondimensional quantity):
$\mathrm{R}_0 \mydef \beta/(\nu+\mu)$,
that expresses the number of individuals
in a naive population that
a carrier of the disease is expected to infect before recovery.
Naturally, we need $\mathrm{R}_0 > 1 $ for long lasting epidemics.
We illustrate the method first for a hypothetical SIRS disease,
with parameters:
$\nu = 10$ [1/year], $\mu = 1/60$ [1/year],
$\gamma = 1.5$ [1/year],
$p=1.2$ [1/year],
and $\mathrm{R}_0 = 10$. The value for the mutation rate was
chosen somewhat high so that relatively smooth traveling pulses
would arise at moderate population sizes.
The random number generator used for the simulations
was a minimal standard generator;
see Press \etal \cite{Press}.
We used the generator of binomial deviates
found in the same reference.

{\bf Single strain.} The behavior of the basic SIR  ($\gamma = 0$) and
SIRS ($\gamma > 0$) model, assuming no mutation or reinfections
at the fluctuation-free limit ($N \to \infty$), can be
written in the form of three ordinary differential equations
for the quantities $S,I$ and $R \in \mathbb{R}^+_0$ 
(see for instance Ref.~\cite{Nasell99a}).
These equations have up to two fixed points,
the disease-free state (extinction of the disease):
$$
S/N = s^0 \mydef 1, ~I/N = i^0 \mydef 0, ~R/N = r^0 \mydef 0,
$$
and the epidemic equilibrium:
\begin{eqnarray*}
S/N = s^\ast \mydef 1/\mathrm{R}_0, \\
I/N = i^\ast \mydef (\mu+\gamma)/(\nu+\mu+\gamma) \left( 1-1/\mathrm{R}_0 \right), \\
R/N = r^\ast \mydef \nu/(\nu+\mu+\gamma) \left( 1-1/\mathrm{R}_0 \right).
\end{eqnarray*}
No epidemic equilibrium exists for $\mathrm{R}_0 <1$:
the disease-free state is a stable fixed point. 
If $\mathrm{R}_0 > 1$,
the epidemic equilibrium is the only stable fixed point.

In a finite population ($N < \infty$) model there is a finite probability of
reaching the extinction of the population state, $S=I=R=0$, at any given time;
this state is absorbing.
There is also an absorbing {\it manifold} of disease-free states $I=0$,
reached after the recovery or death of the last infected individual. 
These two states play a vital role in determining
the truly {\it long-term} expected behavior of the discrete system;
nevertheless, we will work in parameter regimes, and with
population sizes for which {\it over the time of our
simulation/observation} the probability
of such extinctions is negligibly small.
While no direct equivalent to the deterministic epidemic equilibrium
exists in this case, most initial conditions will first approach
a region where
$S/N \approx s^\ast, I/N \approx i^\ast,
R/N \approx r^\ast, $ spend some time there (proportional to $N$),
then decay to the disease-free manifold and eventually
get absorbed by the extinction state.
Our observations can therefore be thought of as ``medium time" 
(as opposed to truly long-time) expected behavior.
When the population size becomes smaller, and the probability of
extinction over a finite observation horizon grows, models based on the
dynamics of rare events become more appropriate than the ones we use here.
See Refs.~\cite{Nasell99a,Nasell02} for very conservative estimates
of the extinction times for the disease.
In the absence of mutations, all epidemics are therefore short lived. 
%

{\bf Multiple strains.} In the presence of mutations and reinfections, a carrier
of a mutant virus in a new class will start to infect
naive individuals, along with some individuals who are not
immune enough to resist the new variant. 
The survival of the disease
hinges on its ability to generate new variants before most of the
population becomes immune.
Infectives are able to start epidemics in other variants,
transmitting the disease to susceptible and recovered individuals with
partial immunity, keeping high rates of infection for
long periods of time.

To study the effect of mutations, we
consider a one-dimensional lattice $i \in \mathbb{Z}$ for the different
strains and allow the reinfection of recovered individuals.
We use the model described in Table~\ref{processes}
with the cross-immunity kernel from Ref.~\cite{LACL}:
$$
K_{i,j} = K(i-j), ~\hbox{with}~ K(\delta)=0 ~\hbox{if}~ \delta<0,
~\hbox{and}~ K(\delta)=\frac{\delta}{(\delta+\delta_0)} ~\hbox{otherwise}.
$$
In this work we are using $\delta_0 = 5$.
This kernel is clearly asymmetric and induces motion to the right
(direction of increasing strain number), 
but respects discrete translational invariance.

Extensive simulations for this stochastic model typically 
reveal --depending on initial distribution, parameters and
cross-immunity kernel--
either a spreading out to zero or {\it apparent traveling pulses}.

\begin{figure}[t!]
\begin{center}
\resizebox{10cm}{!}{
\includegraphics[angle=270]{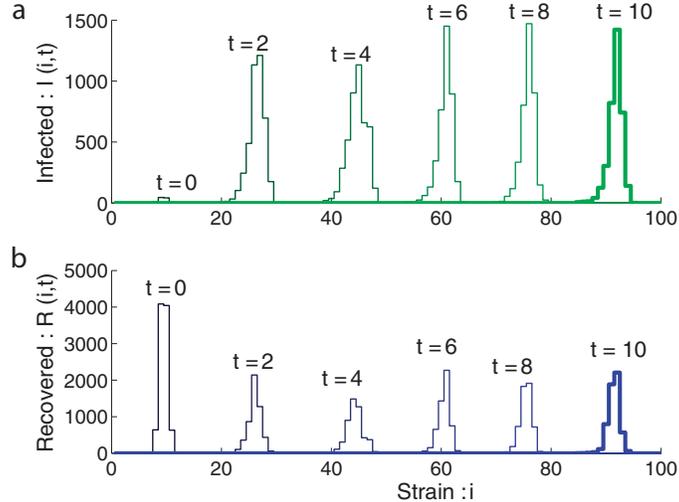}}
\end{center}
\caption[]{
Evolution of the Distribution Functions (DF) for a population of $N=10000$ individuals
over a year.
Parameter values are $\mathrm{R}_0=10, \nu=60 \units{[1/year]},
 \mu=0.0167 \units{[1/year]}, \gamma=1.5 \units{[1/year]}$
and $p=1.2 \units{[1/year]}$.
(a) and (b) show the DF for the infected and the recovered populations.
}
\label{goingtotheright}
\end{figure}

Fig.~\ref{goingtotheright} shows snapshots of a realization
of such a representative pulse. 
After a transient period, the sequence
of infections, mutations, recoveries and reinfections
produces a drift in strain space.
Both distribution functions move in the same direction,
with similar widths but different amplitudes (total number of susceptible
and infected individuals are not the same).
Estimates of the traveling speed in the doubly continuum
(population and strain space) 
version of the model but for different parameters, namely $\gamma=0$,
is found in Ref.~\cite{LACL},
while Brunet and Derrida \cite{BD97,BD99} discusses the 
effect of finite population size in certain individual-based
models with traveling wave behavior.
We will study the dependence of the {\it effective pulse speed} on $N$ in our 
``doubly discrete" model below (a feature reminiscent of other 
``traveling pulses" in lattices, see Refs.~\cite{Kessler1,Kessler2}).

\section{Methods}

\subsection{Coarse Description: the Inverse Cumulative Distribution Function}

We seek continuum ``effective pulses" for our doubly-discrete model.
Representative simulations show that these ``pulses" have relatively
localized support in strain space 
(for most of our simulations they are approximately 5-7 strains wide)
and preserve their coarse shape (modulo random fluctuations) as they move.
Additional fluctuations arise from the discreteness of the lattice,
which only possesses {\it discrete} translational invariance; the
study of effective continuum equations for spatially discrete systems
(lattices, like this one) is the subject of extensive research;
both analytical (\eg \cite{Philip1,Philip2,Philip3,Philip4,Philip5,PadeYannis}) and
equation-free approximations \cite{DiscreteYannis} are possible.
In what follows we will use averaging over several realizations of the
microscopic simulation as well as spatial interpolation (filtering) to
study the system dynamics in terms of {\it effective continuum} observables.
The effective pulses are slightly ``slanted'' to the right, and
this feature may get more pronounced if the mutation rate
$p$ is large or if the number of individuals $N$ is small.
The \emph{distribution functions} (DF's) for the infected and recovered
individuals are the number of individuals, $f(i)$, in each
class, $i$.
Low-dimensional approximations of the distribution
functions seem like natural macroscopic observation variables; we have
chosen instead, and successfully used, low-dimensional 
representations of the
Inverse Cumulative Distribution Function, 
ICDF, also known as percentage point function (PPF):
$$
\begin{array}{c}
g(n) = i ~\hbox{such that} \\
g(n) \le g(m) ~\hbox{if}~n<m
\end{array}
~~\Longleftrightarrow
~~\Sigma_{j=0}^i f(j) = n ~.
$$

Intuitively $g(n)$ corresponds to the {\it location} (\ie strain number)
of individual $n$, after labeling all the individuals from left
to right in strain space.
When $f(i)$ appears as a
localized pulse, $g(n)$ appears as the inverse of a sigmoid function.
%

The ICDF is non-decreasing, and computations with its low
dimensional representation (\eg coarse
projective integration, or fixed point estimation) should not 
destroy its monotonicity; resorting particle labels can restore
monotonicity if necessary.
For a finite number of individuals, $g(i)$ maps a bounded interval 
into a bounded interval (in our convention $g(\cdot)$ maps the interval $[0,1]$
instead of the integers $1 \ldots N$).
The inverse CDF frees us from
adjusting its support as the simulation evolves.
The drift toward the direction of increasing $i$ is seen as an
increase in its \emph{average}, as its
shape remains constant.

\begin{figure}[t!]
\begin{center}
\resizebox{10cm}{!}{
\includegraphics[angle=270]{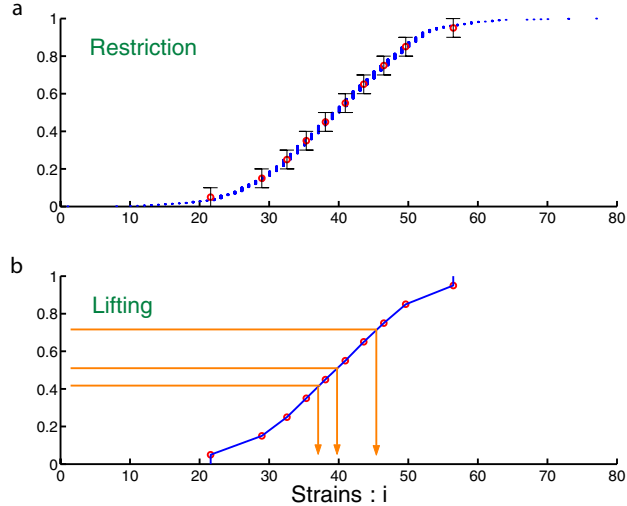}}
\end{center}
\caption[]{
Restriction and lifting scheme:
(a) Using the ICDF representation, we divide the $[0,1]$
interval in 10 subintervals, separate the individuals accordingly,
and compute the average strain of each subgroup, indicated with small circles.
These averages are our low-dimensional representation of the
distribution. We use this reduction with
infected and recovered individuals separately.
(b) Using the averages on each subinterval, we can construct
a distribution, using local linear interpolation,
and generating random numbers uniformly distributed in the
interval $[0,1]$ (vertical coordinate). The images of these
points (in the horizontal coordinate) after being mapped with
the piecewise linear function
provide a distribution of individuals consistent with the
coarse description of the distribution.
}
\label{interpolation}
\end{figure}

Our low-dimensional representation of the coarse system state
using the ICDF is made up of three components:
the number of individuals in the classes $\CS, \CI$ and $\CR$
expressed as ratios with respect to a \emph{reference} number of individuals
$N_\mathrm{ref}$,
the discrete ICDF for the $\CI$ class $g_\CI (x)$,
and the discrete ICDF for the $\CR$ class $g_\CR (x)$.
The vector describing the state of the whole population is then: 
$$
\{ \frac{S}{N_\mathrm{ref}},
\frac{I}{N_\mathrm{ref}},
\frac{R}{N_\mathrm{ref}},
~g_\CI (x_j)_{j=1\ldots M}~,
~g_\CR (x_j)_{j=1\ldots M}~
\} ~.
$$
Using 10 points to discretize the interval $[0,1]$,
our description for the coarse state involves
23 quantities.
A graphical description of the scheme is presented
in Fig.~\ref{interpolation}.

\subsection{The coarse time-stepper}

We symbolize by $\Phi_\tau (\cdot)$ the following sequence
of operations over a point in the low dimensional representation $y$:
(a) lifting (initialization of a large number of realizations $(S,I_i,R_i)$,
each one consistent with $y$); (b) stochastic integration of each independent
realization for a time $\tau$; (c) restriction and averaging over realizations.
Because the problem exhibits traveling behavior, it is convenient to observe
the simulation in a {\it co-moving frame}; in effect, this corresponds to a
{\it shift} of the solution in (strain)-space.
Techniques for determining and effectively implementing this shift have been
described in the literature \cite{Clancy,RTK02,ChenGoldenfeld,DiscreteYannis};
the amount of shift gives then an estimation of the traveling speed.
Steps (a-c) are followed by such a (d) shift (in strain space);
the result is a new ``coarse point" $\Phi_\tau(y)$.
Averaging over a large number of independent realizations
reduces the fluctuations, even if each realization
consists of a finite number of individuals.
Techniques for variance reduction \cite{MO95,OtherGuy} can, in principle,
also be used in evaluating the coarse time-stepper.
The main assumption underlying our computational approach is that the
coarse time-stepper {\it effectively commutes} with the integration
(for time $\tau$) of the unavailable macroscopic equations for the
expected behavior of the epidemic.
Evaluating the coarse time-stepper through short bursts of appropriately
initialized individual-based simulation allows us, as we will see below,
to circumvent the derivation of explicit population-level evolution equations.


\section{Results}

\begin{figure}[t!]
\begin{center}
\resizebox{10cm}{!}{
\includegraphics[angle=270]{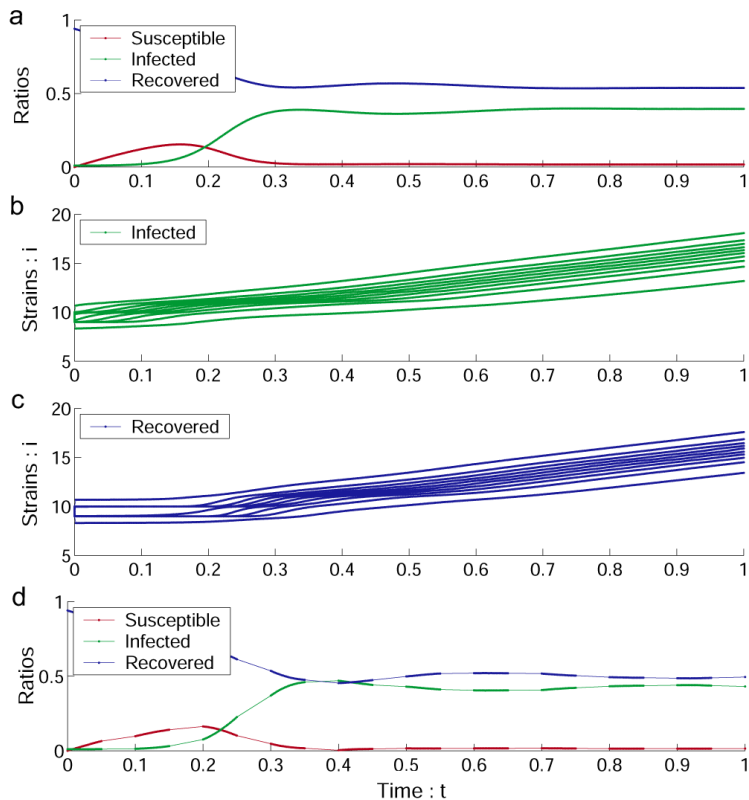}
} 
\end{center}
\caption[]{
(a-c) Evolution of the average over 250 independent
copies of the system,
each one containing around 10000 individuals.
Parameter values are same as in Fig.~\ref{goingtotheright}.
At a given time, the plot
shows the value of 23 modes: 3 in (a), 10 in (b), and
10 in (c). Notice the linear growth of the last 20 modes,
showing a uniform drift in strain space.
These modes capture the shape of the ICDF's
of the $I_i$ and $R_i$ variables, like contour curves. 
As time progresses, shapes of the pulses approach an invariant profile,
so the lines in the two lower graphs get parallel.
(d) Evolution in time using projective forward integration.
The method uses an estimation of the time derivative in the
low dimensional representation, computed after a short run of the
stochastic time-stepper; the projective ``jumps" are marked by thin
lines.
}
\label{parallel}
\end{figure}

We saw that for a single realization, even with large populations,
the amplitudes of the pulses fluctuate around some mean shapes/values,
and the drift in strain space
is often characterized by ``jerky"  behavior (Fig.~\ref{goingtotheright}).
Averaging over many realizations starting with the
same coarse initial condition, we can drastically reduce the variance of
these fluctuations, and approximate the evolution of an effective pulse.
Fig.~\ref{parallel} shows the evolution of a particular initial 
condition associated with a single initial strain; many copies (here 250)
of this initial distribution are evolved in time, with different random seeds.
As the corresponding ICDF's evolve (by spreading and drifting) they
all ultimately approach the same sigmoidal translating profile. 
The (more or less) parallel lines are representative of the drift speed,
and the (approximately one year long) transients are characteristic of
the system's approach to its expected stationary drifting state.

\begin{figure}[t!]
\begin{center}
\resizebox{10cm}{!}{
\includegraphics[angle=270]{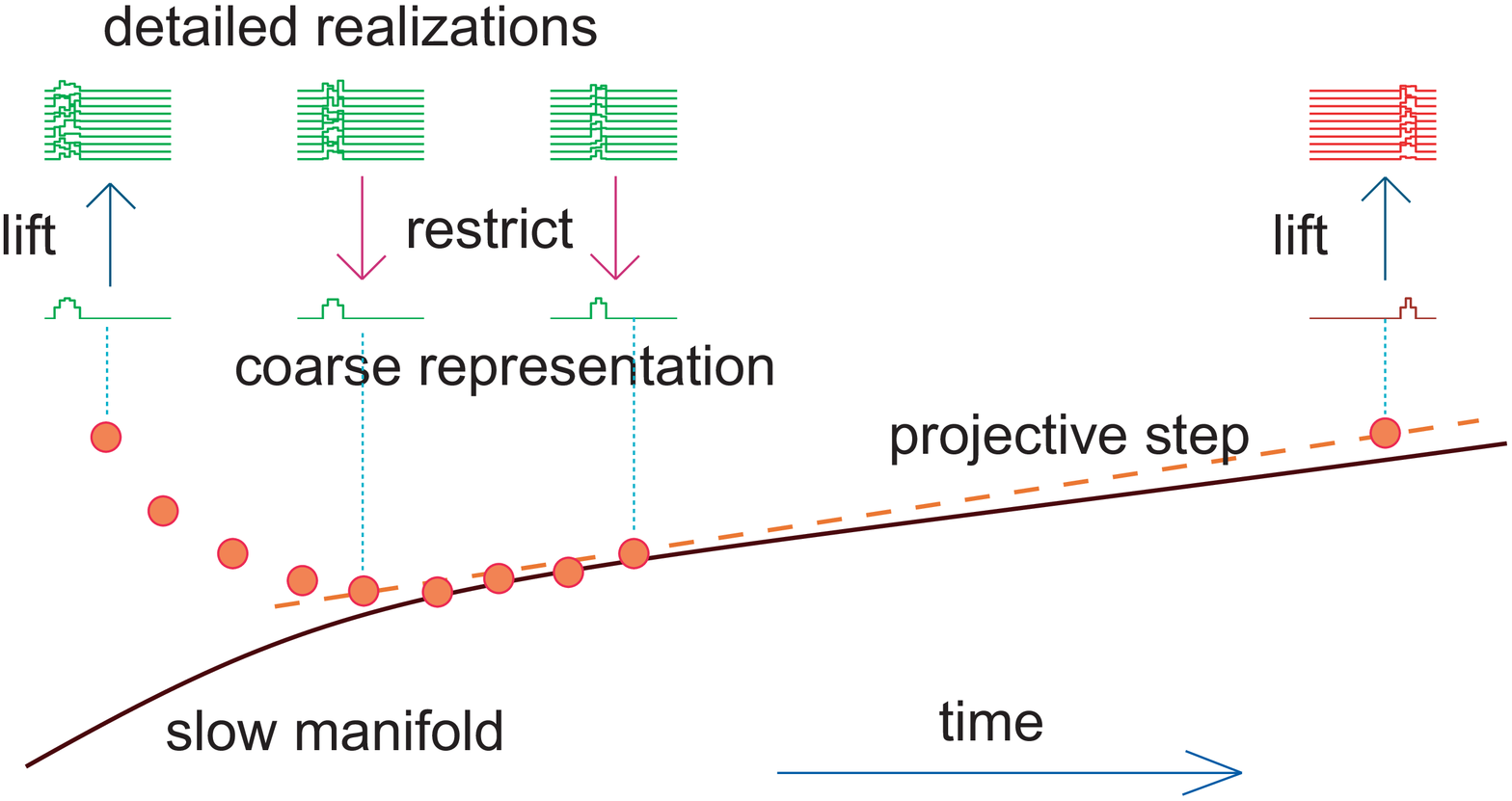}} 
\end{center}
\caption[]{Schematic representation of explicit projective
integration. Translation between coarse representations
and detailed realizations is implemented through lifting and restriction operators.
An initial ``healing" period is required before the detailed
simulation approaches the slow manifold and the coarse time derivative
is estimated, followed by a projection forward in time. 
A single (Forward Euler) projective step is shown here.}
\label{projectivescheme}
\end{figure}

\subsection{Coarse Projective Integration}

Several realizations of the short-time stochastic evolution of an
initial population with certain (coarse) characteristics
can be used to {\it estimate} time derivatives of the 
expected evolution of these coarse characteristics.
As discussed in detail in \cite{Manifesto} the duration of these
stochastic simulation ``bursts" has to be long enough for the
detailed features of the ICDF's to become functionals of (get 
slaved to) the coarse features.
Indeed, the basic feature underlying a macroscopic model is a 
separation of time scales between the low-order (``governing",
or ``master") moments of a stochastically evolving distribution
and the faster, ``slave" higher moments.
The singularly perturbed nature of the dynamics becomes manifest
as higher moments quickly become functionals of the
lower order ones, approaching a low-dimensional, attracting, 
invariant manifold in moments space.
This manifold embodies, in effect, the closure giving rise to 
the lower-dimensional description.
Variance reduction (\eg through the number of realizations) as
well as the duration of the stochastic simulation affects the
quality of the time-derivative estimation.
Using estimates of the time derivatives, we can
implement explicit coarse integration;  we call
this approach `projective forward integration' \cite{GK03, Gear01};
more general extrapolation methods can also be used.
This idea allows one to integrate the detailed problem
only for a short period of time (healing time),
extract the slow evolution of coarse features and then
project their future state.

The procedure is illustrated in Fig.~\ref{projectivescheme}.
The lifting and restriction steps link the coarse
and detailed descriptions of a state.
The {\em projective} step is performed in the coarse
description of the problem, followed by a new lifting step;
the procedure is then repeated.
Notice how a projective step may cause the solution to deviate
from the ``slow manifold", on which higher order moments
are functional of the lower order, governing moments.
A short dynamic simulation {\it constrained} on the coarse
projected state can be used as a preparation step, bringing
the initialization closer to the slow manifold before
letting it free (this is analogous
to constrained molecular dynamics algorithms
like SHAKE \cite{Ryckaert,Carter}).


The method is especially successful in accelerating the
simulation if a significant gap exists between the characteristic
times of a few, slow
modes (corresponding to the macroscopic, expected dynamics) 
and many fast ones (corresponding to the ``slaving" or 
``healing" of the higher moments).
Healing times here are of the order of $0.01$ years
(depending on the quality of the macroscopic description
and the number of independent copies), and
characteristic times for the low-dimensional modes
are of the order of $0.1$ years.

In Fig.~\ref{parallel}d the results of a sequence of projective integrations
are presented for comparison with the `exact' integration
(Fig.~\ref{parallel}a). Thick lines correspond to detailed integrations,
and thin lines to projections based on estimates of time-derivatives.
As we can appreciate, the time horizon used for the detailed integrations
($\tau = 0.05$ years) was long enough to allow healing, but small
enough to capture the meaningful dynamics of the coarse variables,
in particular the transients in the first $0.5$ years.

\subsection{Coarse Fixed-point Computations}

For translationally invariant problems with traveling solutions, 
transforming to a co-traveling frame renders these solutions steady.
This transformation can be performed dynamically,
without {\it a priori} knowledge of the right speed.
Constantly shifting the solution back in (strain)
space through a template-based so-called ``pinning condition" will turn a 
traveling solution into a stationary one; in effect, we are observing
the computation in a co-traveling frame.
In the absence of explicitly closed evolution equations
this procedure can be implemented in discrete time \cite{RTK02}.
Traveling pulses thus become fixed points of the
(dynamically re-shifted) coarse time-stepper:
$$
y - \Phi_\tau (y) = 0.
$$
In additional to the discreteness of the number of the
individuals in a population (smoothened through averaging
many realizations of the process), our strain space is 
discrete, and the solutions can then only have discrete translational invariance
(shift by a single strain along the lattice).
A traveling wave in a spatially discrete problem will then appear, in
a traveling frame, as a (small amplitude) oscillation: it takes a finite
amount of time for the entire shape to progress by exactly one lattice spacing.
When this discreteness-induced modulation is relatively small (as is the case here)
it is still possible, as we mentioned above, (see \cite{DiscreteYannis} for a detailed discussion)
to construct a coarse time-stepper for an {\it effective continuum}
equation.
This is
accomplished once more by averaging over several, shifted, {\it discrete space}
realizations of a spatially continuum initial shape.
It is such {\it effectively translationally invariant} pulses that 
we compute here.

The coarse time-stepper takes an initial condition $y$ in 
the low-dimensional, continuum state representation, 
``lifts" to microscopic realizations consistent with it,
evolves using the stochastic simulator over time 
$\tau$ based on the individual events such as
mutations and infections,
and restricts the averaged results onto the same low-dimensional space.
This yields a map $\Phi_\tau$:
$$
y \mapsto \Phi_\tau (y) = \bar{\Phi}_\tau (y) + \xi ~,
$$
where $\xi$ is some noise that persisted even after
taking the average over many (here 10-250) realizations.
If the system has coarse stationary states,
then we should be able to estimate them
from the statistics of the map $\Phi_\tau$, in particular from
a number of pairs $(y,\Phi_\tau(y))$.

A variety of algorithms can be used to find the fixed point $z$ of
the coarse timestepper. In particular, 
close to this fixed point $z$, using the Jacobian $J$ of $\Phi$ at $z$
we have:
$$
y_1 = z + J \cdot (y_0-z) + \xi + o(y_0-z) ~,
$$
If we have many realizations (through Monte Carlo simulation)
$\{(y_{0,i}, y_{1,i} = \Phi(y_{0,i}))\}_{i=1\ldots N}$
we can estimate $z$ and $J$ from the minimization of
the residual:
$$
\Sigma_{i=1}^N || y_{0,i} - J y_{1,i} - J z ||^2_2 ~.
$$
Setting $w \mydef J z$, the estimation can be done solving
two linear systems, first for $J$ and $w$, then for $z$.
Given the noisy nature of the problem, the Singular Value Decomposition (SVD) 
is used to
invert the relevant matrices in the appropriate subspaces (usually
two or at most three dimensional). 
An estimation of the full
Jacobian would require integrations of the system with
many different initializations.
This method, ``wrapped" as an iterative computational superstructure
around the microscopic simulator, may accelerate the convergence to stable
stationary solutions, and is capable of finding {\it unstable} such
solutions.

In a more general context, matrix-free methods for solving nonlinear equations
(\eg Newton-Krylov methods using GMRES \cite{Kelley99}, based on the coarse time-stepper) 
can be wrapped around the coarse time-stepper to
locate stationary solutions of the unavailable coarse equations.
The Recursive Projection Method of Shroff and Keller \cite{RPM} is one such
algorithm that we have extensively used, and from which our original inspiration
developed.
Such algorithms enable a direct simulation code to perform tasks (such as fixed point
computation) that they have not been in principle designed for.
It is reasonably straightforward to extend the scope of such a computational-enabling
technology to allow direct simulators to perform continuation, bifurcation and
stability analysis, but also controller design and optimization tasks computationally.
Upon convergence of the fixed point algorithms, in particular, matrix-free
eigensolvers (subspace iteration methods, Arnoldi methods based on the time-stepper) can be
used to extract estimates of the eigenvalues of the linearization of the unavailable,
population level equation - thus quantifying coarse stability.


\begin{figure}[t!]
\begin{center}
\resizebox{10cm}{!}{
\includegraphics[angle=270]{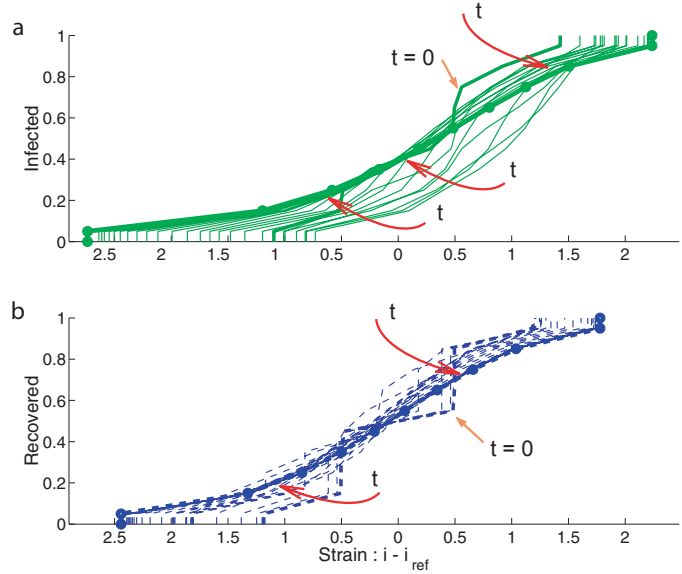}}
\end{center}
\caption[]{
Evolution of the ICDF in the coevolving frame.
(a) and (b) correspond to the infected and recovered populations
respectively.
Distributions at different times are superimposed; the
initial distributions are marked.
As time progresses, distribution shapes converge to the thick dashed lines,
obtained from the fixed points of the coarse representation.
Auxiliary arrows show the progression towards convergence.
Parameter values as in Fig.~\ref{goingtotheright}.
}
\label{sigmoidals}
\end{figure}

\subsection{Continuation of Coarse Solution Branches}

We now present two exploratory sets of computations
for the dependence of the speed of the effective pulse on two
parameters: the strength of infection, quantified by
the non-dimensional number $\mathrm{R}_0$, and the (average) number of
individuals in the population, $N$.
The (rather low) accuracy in both figures shown can be improved using larger
number of realizations every time the coarse time-stepper is used.
This can be accomplished through massively parallel computations, where each processor 
computes a different realization of the same coarse initial condition
for a short time.
Here only 250 realizations were used, 
integrated over a (horizon) time $\tau=0.5$ years.
We used a simple version of continuation: for every new value of the parameter,
the predictor of the new stationary state was obtained through
linear extrapolation from two previously computed fixed points:
$$
y^\mathrm{(0)} (p_{k+1}) =
y (p_k) + \frac{(p-p_k)}{(p_k-p_{k-1})} (y(p_k)-y(p_{k-1})) ~,
$$
and the corrector in the computation of the fixed point
$y^{(\infty)} (p_{k+1}) = y$ 
was the iterative procedure described above.

\begin{figure}[t!] 
\begin{center}
\begin{tabular}{rl}
\resizebox{8.5cm}{!}{
\includegraphics[angle=270]{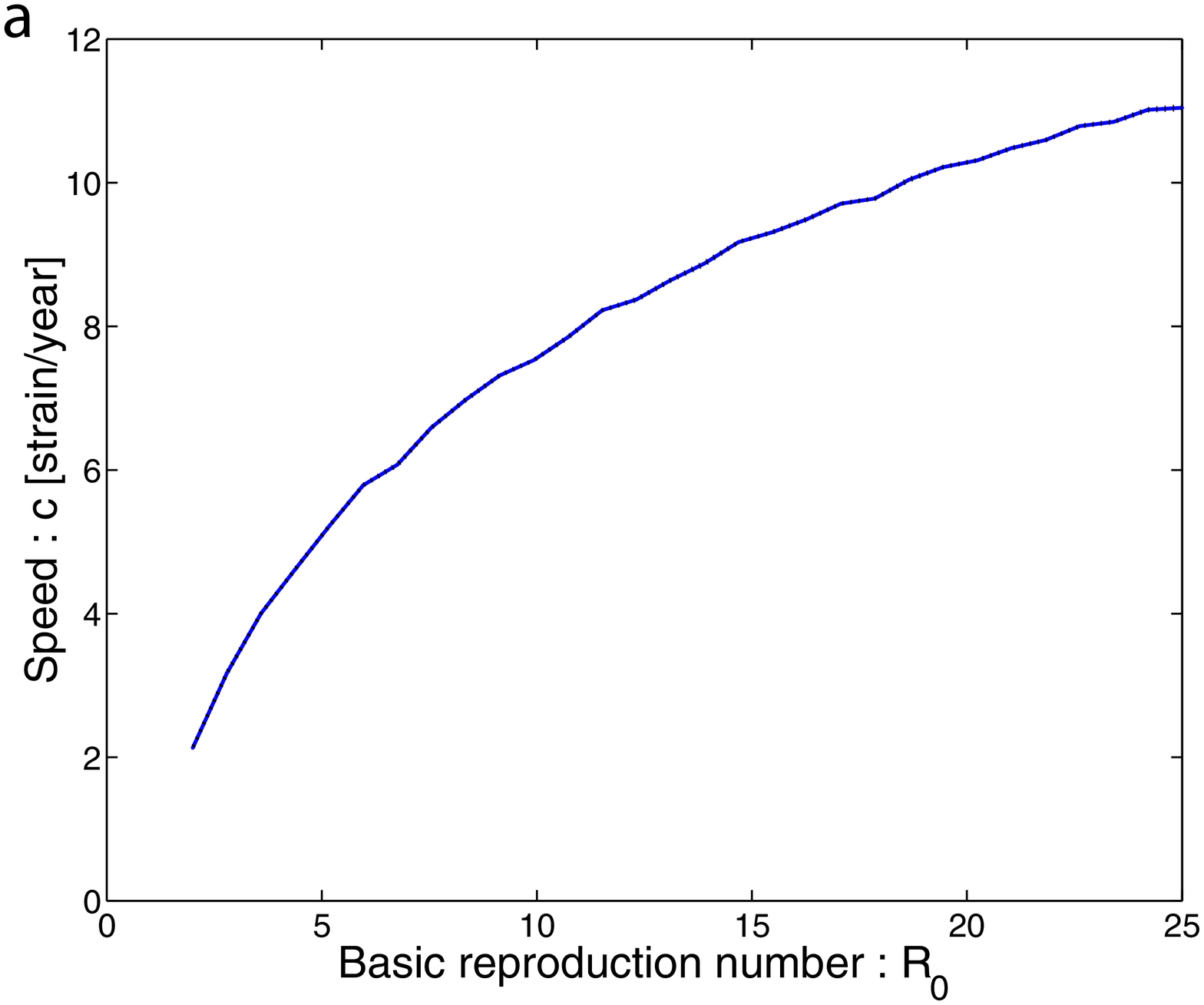}} &
\resizebox{8.5cm}{!}{
\includegraphics[angle=270]{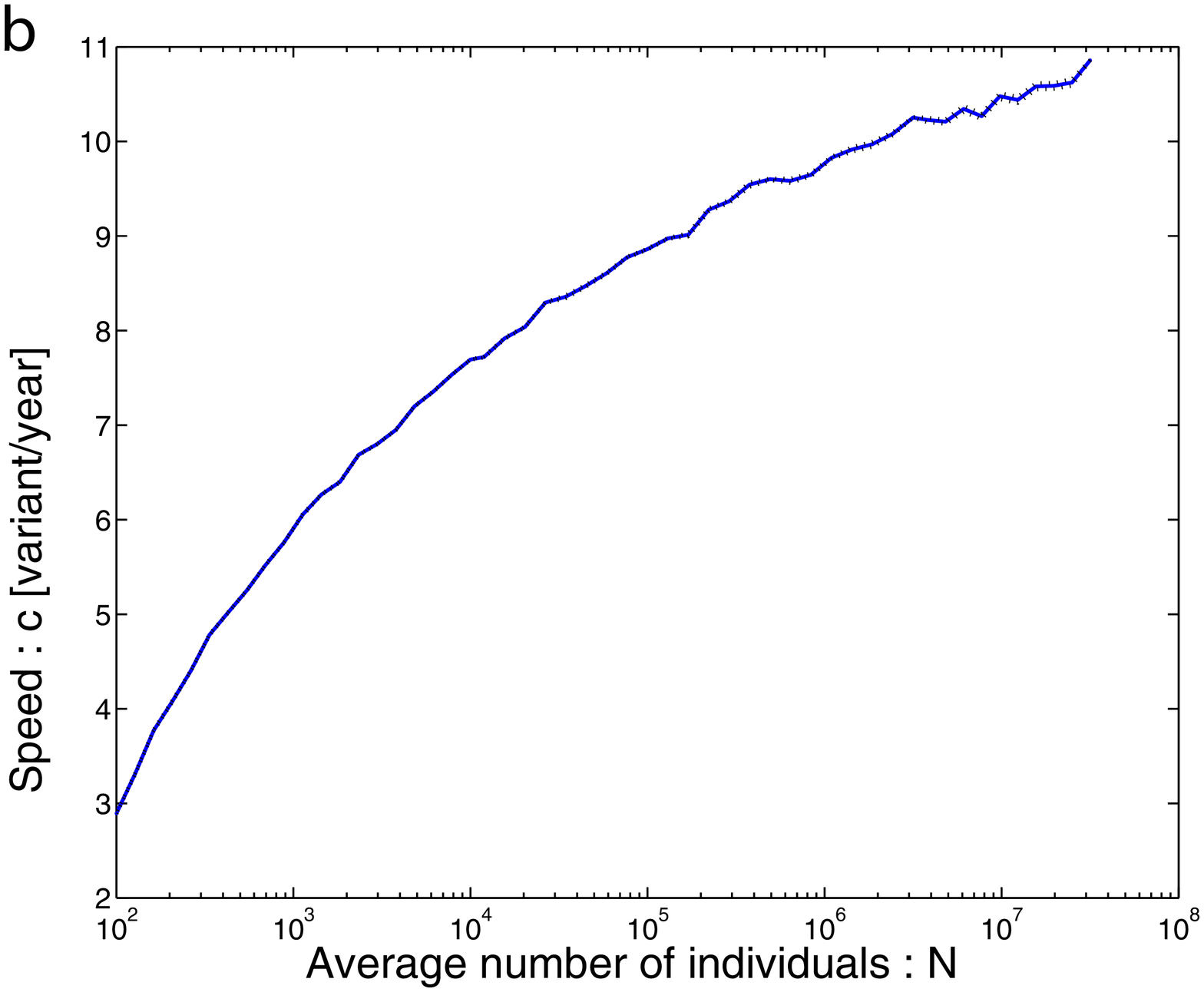}}
\end{tabular}
\end{center}
\caption[]{
(a) Computed speed of the traveling pulses
as we change the basic reproduction number $\mathrm{R}_0$.
The average number of individuals was $10000$.
(b) Computed speed of the traveling pulses
as we change the (average) number of individuals.
We used $\mathrm{R}_0=10$.
Other parameters as in Fig.~\ref{goingtotheright}.
}
\label{cont}
\end{figure}

We can observe in Fig.~\ref{cont}a how
the speed depends on the number $\mathrm{R}_0$,
suggesting a relation of the form $c \propto \sqrt{\mathrm{R}_0 - 1}$.
The existence of a critical value for $\mathrm{R}_0$ can be
rationalized using the one-strain model at the fluctuation-free,
infinite population limit.
If $\mathrm{R}_0 \gtrsim 1$
then there is an epidemic equilibrium with $I \ll N$ and very
low chances of developing a mutant. But if a single mutant appears,
he/she is going to draw the whole population to the new strain.
In Fig.~\ref{cont}b, the speed increases with the number
of individuals, a fact that can also be explained intuitively,
as the probability of having an individual mutation (the `bottleneck')
increases with
the size of the population. Now the curve is far away from
converging, even for $N \sim 10^8$ individuals. The similarity
with $c - c_\infty \propto 1/(\log N)^2$ suggests that some of the ideas of
Ref.~\cite{BD97,BD99} may apply also in this epidemiological model:
the discreteness of the population affects in particular the
tails of the distribution where some bins have $0$ or $1$ individuals.


\section{Summary and Outlook}

We have illustrated the use of equation-free, coarse time-stepper
based techniques, enabling the macroscopic, system level study of
stochastic, individual-based evolutionary epidemiology models.
The technique circumvents the derivation of macroscopic,
population level equations
for the expected behavior of the system; it uses short bursts of
appropriately initialized ensembles of individual-based simulation to
{\it estimate on demand} the quantities necessary for 
system level analysis.
Beyond coarse integration, techniques like fixed point algorithms,
continuation,
and bifurcation / stability analysis are thus enabled (on a light note, 
consider that ``Newton" iteration is performed to find the fixed points
of ``Darwinian" evolution models).

The approach holds promise in studying emergent behavior in
individual-based or agent-based models across several disciplines.
Parametric analysis can be performed directly in terms of microscopic
interaction parameters, as well as of parameters such as the population size.
Various extensions of the approach that have been discussed in other
contexts (see for instance Ref.~\cite{Manifesto})
can also be brought to bear in epidemiological modeling:
the so-called {\it gap-tooth scheme} and {\it patch dynamics} 
have the potential to accelerate
simulations of quantities smoothly distributed over several variables.
Techniques based on matrix-free iterative eigenanalysis can also be used to 
explore when (in parameter space) and how a closure fails, and to establish
whether alternative coarse descriptions of the system
(\eg with more coarse variables) may be useful.

\begin{figure}[t!]
\begin{center}
\includegraphics[width=10cm,angle=270]{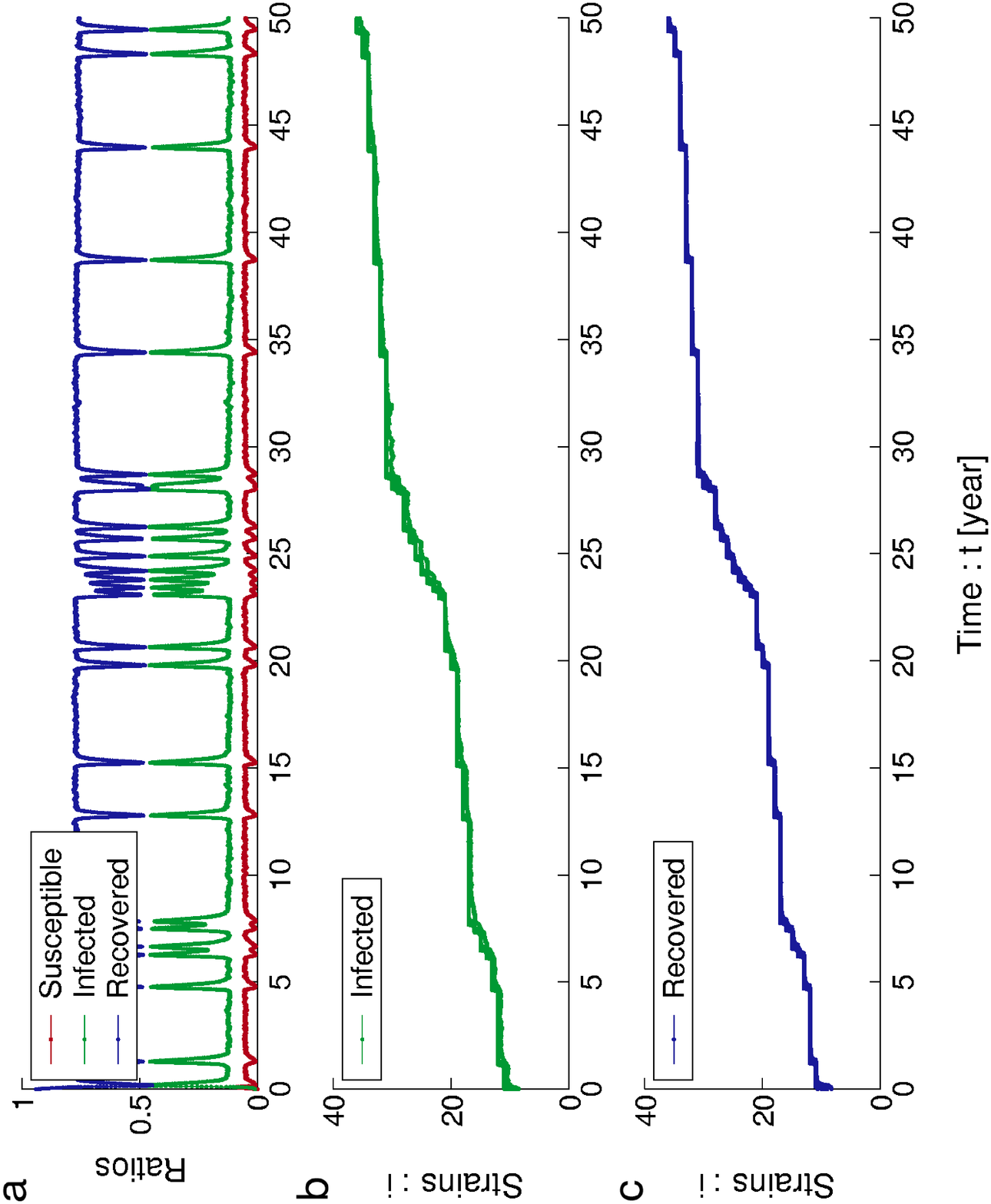}
\end{center}
\caption[]{
Time evolution for a low value of the mutation rate
$p=0.00012 \units{[1/year]}$.
Other parameters as in Fig.~\ref{goingtotheright}.
(a) Ratios $S/N, I/N$ and $R/N$. The number of infected
and susceptible individuals are close to zero most of the time,
except right after a new mutation arises.
(b-c) Approximations of the ICDF's for the infected
and the recovered populations respectively.
Traveling pulses become very narrow (one circulating strain),
and long quiescent periods are punctuated by
sudden jumps that follow the mutation of a single
individual that rapidly infects the rest of the population
with the new strain, followed by fast recruitment and recovery.
}
\label{punctuated_drift}
\end{figure}

Finally, we believe the combination of such coarse techniques
with optimization and the approximation of effective free energy
surfaces may prove useful for the quantitative study of
extinction probabilities and epidemic thresholds;
the simulation in Fig.~\ref{punctuated_drift} is
suggestive of the ``rare event" nature of the epidemic
propagation under such conditions.
For small values of the mutation rate, the pulse itself becomes very
narrow, and the number of infected individuals very small, as most
of the individuals become recovered. 
Drift occurs only if one
infected person mutates to the right $\CI_i \mapsto \CI_{i+1}$
and infects the population of immune individuals to strain $i$
with the new strain $i+1$, producing a shift of one strain to the right.
The speed of these ``punctuated pulses" may be estimated by $c \approx p/2~i^\ast N$:
one half of the mutation rate times the number of individuals in the infected class
at equilibrium; it does not depend on the infectivity.  

Several aspects of equation-free modeling are applicable
in mathematical epidemiology.
The present paper is intended as a proof of principle, outlining the 
computational methodology, and demonstrating the emergence of pulses
--more generally, evolutionary patterns-- in sequence evolution.  
We are currently using this framework to investigate the important case 
of the evolution of the influenza A virus, restricting the SIRS model
to forbid loss of immunity; we intend to extend the approach to a much 
broader class of applications.

\vspace{0.3cm}
\begin{center}{\bf Acknowledgments}\end{center}
\vspace{0.3cm}

{
We are grateful for valuable discussions with Juan Lin,
Viggo Andreasen, Jonathan Dushoff and Gerhard Hummer.
This work was partially supported by the AFSOR, and a NSF-ITR
grant.
}


\addcontentsline{toc}{section}{Bibliography}
\bibliographystyle{plain} 
\bibliography{buc,yannis}

\label{end}\end{document}